\documentclass[aps,prl,preprintnumbers,twocolumn]{revtex4}
\usepackage{graphicx,amsmath}
\usepackage{dcolumn}
\usepackage{amsmath}
\usepackage{amssymb}
\usepackage{amsfonts}
\usepackage{wrapfig}
\usepackage{subfigure}
\def\bra#1{\mathinner{\langle{#1}|}} 
\def\ket#1{\mathinner{|{#1}\rangle}}

\begin{document}

%! program = pdflatex

%\documentclass[12pt,a4paper]{memoir} % for a long document
% See the ``Memoir customise'' template for some common customisations
% Don't forget to read the Memoir manual: memman.pdf

\title{Gibbs Free Energy Analysis of a Quantum Analog of the Classical Binary Symmetric Channel}         % Enter your title between curly braces
\author{David Ford, Physics NPS}        % Enter your name between curly braces
\email[]{dkford@nps.edu}
\affiliation{Department of Physics, Naval Post Graduate School, Monterey, California}
\date{\today}          % Enter your date or \today between curly braces
\begin{abstract}
The Gibbs free energy properties of a quantum {\it send, receive} communications system are studied.
The communications model resembles the classical Ising model of spins on a lattice in that the joint state of the quantum system is the 
product of sender and receiver states. However, the system differs from the classical case in that the sender and receiver spin states are quantum superposition states coupled by a Hamiltonian operator. A basic understanding of these states is directly relevant to communications theory and indirectly relevant to computation since the product states form a basis
(in function space) for entangled states. Highlights of the study include an exact
method for decimation for quantum spins. The main result is that the minimum Gibbs free energy of the
quantum system in the product state is higher (lower capacity) than a classical system with the same parameter values.
The result is both surprising and not. The channel characteristics of the quantum system in the product state are markedly inferior
to those of the classical Ising system. Intuitively, it would seem that capacity should suffer as a result.  
Yet, one would expect entangled states, built from product states, to have better correlation
properties.
\end{abstract}
\maketitle
%%% BEGIN DOCUMENT
\subsection{Introduction}
Potential applications of information theory to computation and communications include optimizations across complex processor/memory hierarchies, within complex biological networks, networks of quantum devices, etc. Often, non-stationarity or random spatial effects make 
analysis even more difficult.  As such is the case, extending relevant engineering principles, such as the concept of channel capacity, to
include  well-established techniques from statistical physics would be beneficial.

In a previous note \cite{fordnote} it was shown that the information theoretic concept of channel capacity is
a particular case of the principle of minimum Gibbs free energy. An analysis based on an extension of the second law
of thermodynamics is found in \cite{kanter}. The system specific work term in the Gibbs formulation was shown to be the ``work of communication''. That is, the work of separating the joint $\{send, receive\}$ signal into its components $\{send\}$ and $\{receive\}$. To make the necessary connections with the physics, the analysis was carried out in the context of the classical, static Ising model of magnetzation \cite{nishimori}, \cite{tanaka}.  The Gibbs free energy properties of a quantum analog of that classical result are investigated below.

In communications theory, the sender of a message is typically responsible for engineering 
the encoding scheme which will efficiently and reliably transfer information to the receiver.
In terms of the variational problem, a search over encoding schemes decides the statistics
of the 1's and 0's of the sent message. To mirror this, the quantum states studied herein are products of 
the sent and received superposition states. In this way, by choosing the sent message superposition state, the sender has chosen the statistics of the measured 1's and 0's of the sent messages. 

The fine grained Hamiltonian operator of the system correlates the the sent and received superposition states. To compute the Gibbs free energy of the system, an exact quantum decimation over the sender and receiver spin sites is performed and coarse grained density operators are computed. In the case of the binary symmetric (i.e. coupled spins with no applied field) channel both the classical and quantum Gibbs free energy minima correspond to maximum single spin (the ``sender's'') entropy. However, it is seen that the Gibbs free energy minimum of the quantum system is higher than that of the analogous classical system.

Two points require some clarification. First, while by adopting this scheme the sender is in control of the statistics of the sent messages, they are not in control of the content of any particular message. Left unchecked this  state of affairs does not directly lead to ``communication'' in the classical sense. However, through a self-measurement of their chosen superposition state, followed by application, if necessary,  of a unitary transformation to the partially collapsed wave function, the sender may effectively communicate with the receiver. This procedure is discussed in the last section in more detail. Secondly, while the states studied
herein are not entangled, they form a basis for the entangled states. A thorough understanding of the properties of the basis states may lead to better computational methods
and offer alternative approaches for the solution of engineering problems.

\subsection{Quantum mechanical background}
Perhaps the closest analog of a classical Ising spin configuration (state) is the direct product of superposition spin states of spin $\frac{1}{2}$ particles. The Pauli representation of the hamiltonian for two coupled spin $\frac{1}{2}$ particles in zero applied field is given by
$$
{\cal{H}}= -J(\sigma_x \cdot \sigma_x + \sigma_y \cdot \sigma_y + \sigma_z \cdot \sigma_z). 
$$
As is traditional, the basis of choice is the one that diagonalizes the spin angular momentum operators $S_z$ and $S^2$ of a single electron
\begin{eqnarray*}
\left(\begin{array}{ccc}
1\\
0\
\end{array}
\right)
=\ket{\uparrow} \,\,\,
\left(\begin{array}{ccc}
0\\
1\
\end{array}
\right)
= \ket{\downarrow}.
\end{eqnarray*}
\noindent In the two electron direct product basis 
\begin{equation}\label{basis}
\{\ket{\uparrow \uparrow}, \ket{\uparrow \downarrow} ,\ket{\downarrow \uparrow} ,\ket{\downarrow \downarrow}\}
\end{equation}  
the hamiltonian takes the matrix form
$$\mathcal{H}= -J 
\left(
\begin{array}{cccc}
1&0&0&0\\
0&-1&2&0\\
0&2&-1&0\\
0&0&0&1\
\end{array}
\right).
$$
\noindent The hamiltonian and density matrix are simultaneously diagonalizable 
\begin{equation}\label{hamrep}
{\cal{H}}= \sum_{i} E_i \ket{i}\!\!\bra{ i }
\end{equation}
\begin{equation}\label{rhorep}
{\cal{\rho}}= \sum_{i} \frac{e^{-\beta E_i}}{Q} \ket{i}\!\!\bra{ i }
\end{equation}
\noindent where $Q = \sum_{i} e^{-\beta E_i}$. In the chosen basis (\ref{basis}), the energy eigenvalues
and eigenstates, $\{E_i$, $\ket{i}\}$, are
$$
\begin{array}{ll}
&\left\{-J, 
\left(\begin{array}{ccc}
&1&0\\
&0&0\
\end{array}
\right)
\right\},
\left\{-J,
\left(\begin{array}{ccc}
&0&0\\
&0&1\
\end{array}
\right)
\right\},\\
\\& 
\left\{ -J,
 \left(\begin{array}{ccc}
&0&\frac{1}{\sqrt{2}}\\
&\frac{1}{\sqrt{2}}&0\
\end{array}
\right)
\right\},
\left\{3J,
\left(\begin{array}{ccc}
&0&\frac{1}{\sqrt{2}}\\
&-\frac{1}{\sqrt{2}}&0\
\end{array}
\right) 
\right\}.
\end{array}
$$
In shorthand notation
\begin{eqnarray*}
{\cal{H}} &=& -J \ket{\uparrow \uparrow} \!\!  \bra{\uparrow \uparrow} -J \ket{\downarrow \downarrow} \!\! \bra{\downarrow \downarrow}\\
& & -J \ket{{\scriptstyle sym}} \!\! \bra{{\scriptstyle sym}}+ 3J \ket{{\scriptstyle anti}}  \!\!\bra{{\scriptstyle anti}}.
\end{eqnarray*}

In the model it is assumed that the sender transmits a state of the form
\begin{equation}\label{sender}
\psi_{send}=\cos \theta_s  \ket{\uparrow} + \sin\theta_s \ket{\downarrow}
\end{equation}
\noindent Similarly the receiver will obtain
\[\psi_{rcv}=\cos \theta_r  \mid\uparrow\rangle + \sin\theta_r \mid\downarrow\rangle\]
The joint {\it communication} state $\psi_{c}$ generated by the send, receive process is 
\begin{equation*}
\psi_{c}= \psi_{send} \otimes \psi_{rcv}.
\end{equation*}
\noindent In the chosen basis (\ref{basis}), the communication state admits the representation
\begin{equation}\label{psirep}
\left(\begin{array}{ccc}
\cos \theta_s \cos \theta_r& \cos \theta_s \sin \theta_r\\
\cos \theta_r  \sin\theta_s& \sin \theta_r  \sin\theta_s\
\end{array}
\right).
\end{equation}
\noindent Note that the communications model lives in a subset of all possible two spin states. It is not the most general possible representation. In particular, (\ref{psirep}) is not an entangled state. However,
entangled states may be well approximated by sums of states of this type.

The action of ${\cal{H}}$ on the communication state $\psi_{c}$ is
\begin{eqnarray*}
\mathcal{H} \ket{\psi_{c}} &=& -J \cos \theta_s \cos \theta_r \, \ket{\uparrow \uparrow}
-J \sin \theta_s \sin \theta_r \, \ket{\downarrow \downarrow}\\ [1ex]
& & -J \, \frac{\sin(\theta_s + \theta_r)}{\sqrt{2}} \, \ket{{\scriptstyle sym}}
+3J \, \frac{\sin(\theta_r  - \theta_s)}{\sqrt{2}} \, \ket{{\scriptstyle anti}} \\
\end{eqnarray*}
\noindent or
\begin{eqnarray*}
 \left(\begin{array}{ccc}
 -J  \cos \theta_s \cos \theta_r  & 
\begin{array}{ll} 
& \frac{3}{2} J \sin(\theta_s-\theta_r) \\ [0.5ex]& -\frac{1}{2}J \sin(\theta_s+\theta_r) 
\end{array}
\\
\\
\begin{array}{ll} 
&  -\frac{3}{2}J \sin(\theta_s-\theta_r) \\ [0.5ex] & \;\;  -\frac{1}{2}J \sin(\theta_s+\theta_r) 
\end{array}
&
-J\sin \theta_s \sin \theta_r
\end{array}
\right).
\end{eqnarray*}
\noindent The average energy of the system under $\cal{H}$, a function of $\theta_s$ and $\theta_r$,
is given by
$$
\bra{\psi_{c}}{\cal{H}}\ket{\psi_{c}}=-J \cos(2(\theta_s-\theta_r)).
$$

\begin{figure}[htbp]
\begin{center}
\leavevmode
\includegraphics[width=60mm,keepaspectratio]{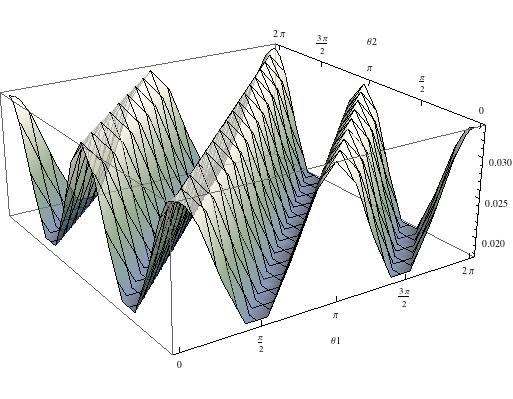}
\caption{ $\bra{\psi_{c}}{\cal{\rho}}\ket{\psi_{c}}$ for the hamiltonian $-J \sigma \cdot \sigma$ with zero applied field.}
\label{wrinkle}
\end{center}
\end{figure}

\noindent The density operator average, $\bra{\psi_{c}}{\cal{\rho}}\ket{\psi_{c}}$, also defines a function of
of $\theta_s$ and $\theta_r$

\begin{eqnarray}\label{prob}
\frac{\mathcal{P}(\theta_s,\theta_r)}{\pi^2} &=&\\ \nonumber
&&
\hspace{-4ex} 
\frac{1}{\pi^2}
\frac{1+3e^{\beta 4J}+(e^{\beta 4J}-1) \cos(2(\theta_s-\theta_r))}{4(1+3e^{\beta 4J})}.
\end{eqnarray}
\noindent This function determines the probability density of the communications state generated by
the send, receive state pairs $\{\theta_s, \theta_r \} \in [0,2\pi)^2$ 
under the influence of the hamiltonian $\mathcal{H}$. The factor of $\pi^2$ in the denominator is the total statistical weight of $\psi_c$
under $\mathcal{H}$. This probability density function is plotted in figure \ref{wrinkle}. The coupling is apparent in the figure. The most likely
states occur when $\theta_s=\theta_r$. The least likely when $\theta_s$ and $\theta_r$ are phase shifted by $\frac{\pi}{2}$.

\subsection{The quantum channel}
In practice, the communication of messages occurs via a set of codewords. A choice of codeword
set fixes the statistics (the sender's marginal distribution) of the sent signal. The received statistics may be different due to transmission errors. In equation (\ref{sender}), the sender's statistics are governed by $\psi_{send}$. That is, by the angle
$\theta_s$. For each $\theta_s$, the hamiltonian of the system determines the joint statistics.
In other words, for each $\theta_s$, the system hamiltonian induces a distribution on the received messages statistics. That is, it induces a distribution on $\theta_{r}$. 

For example, at fixed $\theta_{s}$, the probability of sending an $\uparrow$ and receiving an $\uparrow$ varies with the receiver's $\theta_{r}$. Over a sufficiently long observation period, an average probability characterizing the $\theta_{r}$ variation emerges so that it makes sense to speak of an averaged sender's density operator. It is assumed that this empirical averaging is equal to the thermal average over $\theta_{r}$. 

The eigenstates described so far, for example in equations (\ref{hamrep}) and (\ref{rhorep}), correspond to operators which act on the electron pair. These joint eigenstates may be decomposed
into pieces (herein referred to as {\it eigenpieces}) according to their action on, say, the sender's spin state. Such a decomposition facilitates thermal averaging over $\theta_{r}$. Table \ref{table:proj} gathers together the information necessary to analyze the thermal averages of the receiver and construct the sender's density operator in the presence of an applied field.
\begin{table}[htpb] 
\caption{Decimated Density Operator (sender)} % title of Table 
\centering      % used for centering table 
\begin{tabular}{c c c c}  % centered columns (4 columns) 
\hline\hline  %inserts double horizontal lines 
\\  [-1.5ex]   %inserts vertical space before text starts 
label & weight & projection amplitudes & operator\\ [.5ex] % inserts table heading 
\hline   %inserts single horizontal line 
\\  [-1.5ex]                %inserts vertical space before text starts 
(u) & $e^{- \beta(2h - J)}$ \; & \; $\bra{\psi_{r}}  \! \uparrow \rangle \! \langle \uparrow  \! \ket{\psi_{r}}$ & \; $\ket{\uparrow}\bra{\uparrow}$\\
\\
(d) & $e^{- \beta(-2h - J)}$\; & \; $\bra{\psi_{r}}  \! \downarrow \rangle \! \langle \downarrow  \! \ket{\psi_{r}} $& \; $\ket{\downarrow}\bra{\downarrow}$\\
\\
(s1) & $e^{\beta J}$\; & \; $\frac{1}{2}\bra{\psi_{r}}  \! \downarrow \rangle \! \langle \downarrow  \! \ket{\psi_{r}}$ & \; $\ket{\uparrow}\bra{\uparrow}$\\
\\  [-2.5ex]
(s2) & $e^{\beta J}$\; & \; $\frac{1}{2}\bra{\psi_{r}}  \! \downarrow \rangle \! \langle \uparrow  \! \ket{\psi_{r}}$ & \; $\ket{\uparrow}\bra{\downarrow}$\\
\\  [-2.5ex]
(s3) & $e^{\beta J}$\; & \; $\frac{1}{2}\bra{\psi_{r}}  \! \uparrow \rangle \! \langle \downarrow  \! \ket{\psi_{r}}$ & \; $\ket{\downarrow}\bra{\uparrow}$\\
\\  [-2.5ex]
(s4) & $e^{\beta J}$\; & \; $\frac{1}{2}\bra{\psi_{r}}  \! \uparrow \rangle \! \langle \uparrow  \! \ket{\psi_{r}}$ & \; $\ket{\downarrow}\bra{\downarrow}$\\
\\
(a1) & $e^{-\beta 3J}$\; & \; $\frac{1}{2}\bra{\psi_{r}}  \! \downarrow \rangle \! \langle \downarrow  \! \ket{\psi_{r}}$ & \; $\ket{\uparrow}\bra{\uparrow}$\\
\\  [-2.5ex]
(a2) &$e^{-\beta 3J}$\; & \!\!\!  $-\frac{1}{2}\bra{\psi_{r}}  \! \downarrow \rangle \! \langle \uparrow  \! \ket{\psi_{r}}$ & \; $\ket{\uparrow}\bra{\downarrow}$\\
\\  [-2.5ex]
(a3) & $e^{-\beta 3J}$\; & \!\!\!  $-\frac{1}{2}\bra{\psi_{r}}  \! \uparrow \rangle \! \langle \downarrow  \! \ket{\psi_{r}}$ & \; $\ket{\downarrow}\bra{\uparrow}$\\
\\  [-2.5ex]
(a4) & $e^{-\beta 3J}$\; & \; $\frac{1}{2}\bra{\psi_{r}}  \! \uparrow \rangle \! \langle \uparrow  \! \ket{\psi_{r}}$ & \; $\ket{\downarrow}\bra{\downarrow}$\\ [1ex]       % [1ex] adds vertical space 
\hline     %inserts single line 
\end{tabular} 
\label{table:proj}  % is used to refer this table in the text 
\end{table} 

The row corresponding to eigenpiece (u) of Table \ref{table:proj} indicates how the receiver averaged statistical weight of sending an $\uparrow$ and receiving an $\uparrow$ (as a function of $\theta_{s}$) is obtained. There is a projection of $\psi_{send}$ onto the appropriate eigenpiece and an integration over the $\psi_{rcv}$ amplitudes. The calculation is presented in equation (\ref{upprobs}). Similarly, the receiver averaged statistical weight of sending an $\uparrow$ and receiving a $\downarrow$ (as a function of $\theta_{s}$) is obtained by integrating over $\psi_{rcv}$
amplitudes in eigenpieces (s1) and (a1). That calculation is presented in equation (\ref{udprobs}).
 It should be noted that the mixed terms, eigenpieces (s2), (s3), (a2) and (a3) make no contribution to the receiver averaged operator.

\begin{eqnarray}\label{upprobs}
\rho_{\uparrow \uparrow} &=&\\ \nonumber
& &
\hspace{-4ex}
\frac{1}{Q} \int_0^{2\pi} \frac{{\textstyle d\theta_{r}}}{\pi} {\textstyle \, e^{-\beta(2h- J)} \bra{\psi_{r}}  \! \uparrow \rangle \! \langle \uparrow  \! \ket{\psi_{r}} \,  \bra{\psi_{s}}  \! \uparrow \rangle \! \langle \uparrow  \! \ket{\psi_{s}} }
\\ \nonumber
\\
\label{udprobs}
\rho_{\uparrow \downarrow}&=&\\ \nonumber
& &
\hspace{-4ex}
\frac{1}{Q}\int_0^{2\pi}   \frac{d\theta_{r}}{\pi}{\textstyle \left(\frac{e^{-\beta(- J)} + e^{-\beta3J}}{2}\right)
\! \bra{\psi_{r}}  \! \downarrow \rangle \! \langle \downarrow  \! \ket{\psi_{r}}  \bra{\psi_{s}}  \! \uparrow \rangle \! \langle \uparrow  \! \ket{\psi_{s}} }.
\end{eqnarray}

Observe that in  the calculations above, the integration over $\theta_r$ is uniformly weighted over the region $[0,2\pi]$. However, as has been mentioned, the choice of codeword
set fixes the statistics (the sender's marginal distribution) of the sent signal. To mirror this, the $\theta_s$ integration is performed with a delta function so that only the sender's chosen value, ``$\theta_{\ast}$'' contributes to the sum over states.  The net effect is that there are fewer available states and Q is interpreted as
\begin{eqnarray}\label{newQ}
Q &=& 
\\ \nonumber
&& \hspace{-4ex} \int_0^{2\pi} d \theta_s \delta(\theta_s - \theta_{\ast}) \, \int_0^{2\pi} \frac{d\theta_{r}}{\pi}  \,  \sum_{i} e^{-\beta E_i} \,  \bra{\psi_{c}}  \! i \rangle \! \langle i  \! \ket{\psi_{c}}.
\end{eqnarray}

The probability of receiving an $\uparrow$ conditional upon an $\uparrow$ being sent is computed as 
\begin{eqnarray*}
 P_{\uparrow | \uparrow} &=& \frac{\rho_{\uparrow \uparrow}}{\rho_{\uparrow \uparrow}+\rho_{\uparrow \downarrow}}\\
 &=&\frac{\frac{e^{-\beta(2h-J)} \cos ^2(\theta_{\ast})}{Q}}{\frac{e^{-\beta(2h-J)} \cos ^2(\theta_{\ast})}{Q}+\frac{e^{-\beta J } 
\cos ^2(\theta_{\ast}) \cosh (2 \beta J )}{Q}}
\end{eqnarray*}
\noindent Similarly, the probability of receiving an $\downarrow$ conditional upon an $\uparrow$ being sent is computed as 
\begin{eqnarray*}
 P_{\downarrow | \uparrow}&=&\frac{\rho_{\uparrow \downarrow}}{\rho_{\uparrow \uparrow}+\rho_{\uparrow \downarrow}}\\
&=&\frac{\frac{e^{-\beta J } 
\cos ^2(\theta_{\ast}) \cosh (2 \beta J )}{Q}}{\frac{e^{-\beta(2h-J)} \cos ^2(\theta_{\ast})}{Q}+\frac{e^{-\beta J } 
\cos ^2(\theta_{\ast}) \cosh (2 \beta J )}{Q}}
\end{eqnarray*}

\noindent In this way the elements of the quantum channel 
\begin{eqnarray}\label{chanmat}
\left(
\begin{array}{cc}
 P_{\uparrow | \uparrow} &P_{\downarrow | \uparrow} \\ 
P_{\uparrow | \downarrow} & P_{\downarrow | \downarrow}  
\end{array} 
\right)
\end{eqnarray}
\noindent are determined from the thermal averages.

At zero applied field, $P_{\uparrow | \uparrow}$ and $P_{\downarrow | \downarrow}$  are equal to each other and (\ref{chanmat}) becomes a symmetric channel. Due to the symmetry of the hamiltonian and the fact that the rows of the channel must sum to one, a single entry (say $P_{\uparrow | \uparrow}$) is enough to determine the matrix. Under these conditions (\ref{chanmat}) becomes

\begin{eqnarray}\label{chanmat2}
\left(
\begin{array}{c c c}
\alpha&  &1-\alpha\\
  &  &\\
1-\alpha&  &\alpha
\end{array}
\right)
\end{eqnarray} 
\noindent with 
$$
\alpha=\frac{2 e^{4 J \beta }}{1+3 e^{4 J \beta }}.
$$
The channel properties are independent of $\psi_{send}$, i.e. the angle $\theta_s$. 
This is the reminiscent of the classical case where
the ``noise'' is independent of the properties of the information source. However, it differs from the classical case in that
arbitrarily increasing the coupling strength, $J$, does {\it not} arbitrarily improve the channel probabilities. 

\subsection{Construction of the coarse grained operators }
\subsubsection{The hamiltonian and density operators}
The Pauli representation of the hamiltonian for two coupled spin $\frac{1}{2}$ particles in an applied field  $h$ is given by
\begin{equation}\label{h_hammy}
{\cal{H}}= -J(\sigma_x \cdot \sigma_x + \sigma_y \cdot \sigma_y + \sigma_z \cdot \sigma_z) + h(\sigma_z \cdot \sigma_o + \sigma_o \cdot \sigma_z ).
\end{equation}

\noindent In the energy basis the hamiltonian takes the form
\begin{eqnarray*}
{\cal{H}} &=&
 (2h-J) \ket{\uparrow \uparrow} \!\!  \bra{\uparrow \uparrow} +(-2h-J) \ket{\downarrow \downarrow} \!\! \bra{\downarrow \downarrow}
 \\ \nonumber
& &
\hspace{-0ex}
-J \ket{{\scriptstyle sym}} \!\! \bra{{\scriptstyle sym}}+ 3J \ket{{\scriptstyle anti}}  \!\!\bra{{\scriptstyle anti}}
\end{eqnarray*}
\noindent and the density operator
\begin{eqnarray} \label{hrho} \nonumber
{\rho} &=& \frac{1}{Q} (
\exp^{-\beta(2h-J)} \ket{\uparrow \uparrow} \!\!  \bra{\uparrow \uparrow} + \exp^{-\beta(-2h-J)} \ket{\downarrow \downarrow} \!\! \bra{\downarrow \downarrow}\\
& &
\hspace{-3ex}
+ \exp^{-\beta(-J)} \ket{{\scriptstyle sym}} \!\! \bra{{\scriptstyle sym}}+  \exp^{-\beta3J} \ket{{\scriptstyle anti}}  \!\!\bra{{\scriptstyle anti}}
\; ).
\end{eqnarray}

\noindent Recall from equation (\ref{prob}) that the density operator average, $\bra{\psi_{c}}{\cal{\rho}}\ket{\psi_{c}}$, defines $\mathcal{P}$, a function
of $\theta_s$ and $\theta_r$ with

\begin{equation}\label{prob2}
\int_0^{2\pi}\frac{d \theta_s}{\pi} \left( \int_0^{2\pi}  \frac{d \theta_r}{\pi} \mathcal{P}(\theta_s,\theta_r)\right) =1.
\end{equation}
\noindent The result of inner integration defines a coarsened density $\tilde{\mathcal{P}}(\theta_s)$, the senders marginal.
One may also use the integration over $\theta_r$ to define the quantum decimated hamiltonian operator.
The details necessary for its construction are gathered together in Table \ref{table:decimate2}. Recall that, to mirror the 
effect of the sender being in control of the sent message statistics, the integral over $\theta_s$ is performed with
a delta function. See equation (\ref{newQ}).

\subsubsection{Decimation of the quantum system}

\begin{table}[htpb] 
\caption{Decimated Density Operator} % title of Table 
\centering      % used for centering table 
\begin{tabular}{c c c c}  % centered columns (4 columns) 
\hline\hline  %inserts double horizontal lines 
\\  [-1.5ex]   %inserts vertical space before text starts 
label & weight & projection amplitudes & operator\\ [.5ex] % inserts table heading 
\hline   %inserts single horizontal line 
\\  [-1.5ex]                %inserts vertical space before text starts 
(u) & $e^{- \beta(2h - J)}$ \; & \; $\bra{\psi_{r}}  \! \uparrow \rangle \! \langle \uparrow  \! \ket{\psi_{r}}$ & \; $\ket{\uparrow}\bra{\uparrow}$\\
\\ [-2.5ex]
(s1) & $e^{\beta J}$\; & \; $\frac{1}{2}\bra{\psi_{r}}  \! \downarrow \rangle \! \langle \downarrow  \! \ket{\psi_{r}}$ & \; $\ket{\uparrow}\bra{\uparrow}$\\
\\  [-2.5ex]
(a1) & $e^{-\beta 3J}$\; & \; $\frac{1}{2}\bra{\psi_{r}}  \! \downarrow \rangle \! \langle \downarrow  \! \ket{\psi_{r}}$ & \; $\ket{\uparrow}\bra{\uparrow}$\\
\\ 
(d) & $e^{- \beta(-2h - J)}$\; & \; $\bra{\psi_{r}}  \! \downarrow \rangle \! \langle \downarrow  \! \ket{\psi_{r}} $& \; $\ket{\downarrow}\bra{\downarrow}$\\
\\ [-2.5ex]
(s4) & $e^{\beta J}$\; & \; $\frac{1}{2}\bra{\psi_{r}}  \! \uparrow \rangle \! \langle \uparrow  \! \ket{\psi_{r}}$ & \; $\ket{\downarrow}\bra{\downarrow}$\\
\\ [-2.5ex]
(a4) & $e^{-\beta 3J}$\; & \; $\frac{1}{2}\bra{\psi_{r}}  \! \uparrow \rangle \! \langle \uparrow  \! \ket{\psi_{r}}$ & \; $\ket{\downarrow}\bra{\downarrow}$\\
[1ex]       % [1ex] adds vertical space 
\hline     %inserts single line 
\end{tabular} 
\label{table:decimate2}  % is used to refer this table in the text 
\end{table} 
The mixed term eigenpiece projections, (s2), (a2), (s3), (a3),  in Table \ref{table:proj}  average to zero. The remaining terms are collected together according to single electron
eigenstates in Table \ref{table:decimate2}. 

The form of the reduced hamiltonian operator is given by
$$
\tilde{\mathcal{H}}= \tilde{H}_\uparrow \ket{\uparrow}\!\!\bra{\uparrow} + \tilde{H}_\downarrow \ket{\downarrow}\!\!\bra{\downarrow}.
$$
\noindent The details are readily obtained from Table \ref{table:decimate2} by inspection
\begin{eqnarray*}
\tilde{H}_\uparrow &=&
\hspace{-0ex} 
-\frac{1}{\beta}\log \int_0^{2\pi} 
 \frac{d \theta_r}{\pi} \hspace{1ex}
e^{- \beta(2h - J)} \mid \bra{\psi_{r}}  \! \uparrow \rangle \mid^2 \\
&& + e^{\beta J} \frac{1}{2} \mid\bra{\psi_{r}}  \! \downarrow \rangle\mid^2
+e^{-\beta 3J} \frac{1}{2} \mid \bra{\psi_{r}}  \! \downarrow \rangle\mid^2
\\ [1ex]
&=&-\frac{1}{\beta}\log 
 \left( 
e^{- \beta(2h - J)} + \frac{1}{2}e^{\beta J}  + \frac{1}{2} e^{-\beta 3J} 
\right) 
\end{eqnarray*}

\begin{eqnarray*}
\tilde{H}_\downarrow &=& -\frac{1}{\beta}\log \int_0^{2\pi}  
\frac{d \theta_r}{\pi} \hspace{1ex}
e^{- \beta(-2h - J)} \mid \bra{\psi_{r}}  \! \downarrow \rangle \mid^2 \\
&&
+e^{\beta J} \frac{1}{2} \mid \bra{\psi_{r}}  \! \uparrow \rangle\mid^2
+e^{-\beta 3J} \frac{1}{2}  \mid \bra{\psi_{r}}  \! \uparrow \rangle\mid^2
\\  [1ex]
&=&-\frac{1}{\beta}\log 
 \left( 
e^{- \beta(-2h - J)} + \frac{1}{2}e^{\beta J}  + \frac{1}{2} e^{-\beta 3J} 
\right) 
\end{eqnarray*}

\noindent The form of the reduced density operator is then
\begin{equation}\label{rhohat}
\tilde{\rho} = \frac{e^{-\beta \tilde{H}_\uparrow}}{\tilde{Q}}  \ket{\uparrow}\!\!\bra{\uparrow} + \frac{e^{-\beta \tilde{H}_\downarrow}}{\tilde{Q}} \ket{\downarrow}\!\!\bra{\downarrow}
\end{equation}
\noindent with (recall equation (\ref{newQ})) 
\begin{eqnarray*}
\tilde{Q} &=& \int_0^{2\pi} 
d \theta_s \delta(\theta_s - \theta_{\ast})\\
&&
\hspace{10ex}
\times
\left( 
e^{-\beta \tilde{H}_\uparrow} \mid\bra{\psi_{s}}  \! \uparrow \rangle\mid^2 + e^{-\beta \tilde{H}_\downarrow} \mid\bra{\psi_{s}}  \! \downarrow \rangle\mid^2
 \right) \\
&=& \left( e^{- \beta(2h - J)} + \frac{1}{2} e^{\beta J} + \frac{1}{2} e^{-\beta 3J} \right) \cos^2{\theta_{\ast}}
\\
& &+ \left( e^{- \beta(-2h - J)} + \frac{1}{2} e^{\beta J} + \frac{1}{2} e^{-\beta 3J} \right) \sin^2{\theta_{\ast}}\\
&=& e^{- \beta(2h - J)}  \cos^2{\theta_{\ast}}+ \frac{1}{2} e^{\beta J} \\
& & + \frac{1}{2} e^{-\beta 3J} + e^{- \beta(-2h - J)} \sin^2{\theta_{\ast}}\\
&=& Q.
\end{eqnarray*}

\noindent A similar decimation is carried out over the sender's spin site.  A  delta function ensures that all the probability mass is assigned to the sender's {\it choice} of $\theta_{s} = \theta_{\ast}$. Table \ref{table:decimate2} eigenpieces (u), (s4) and (a4) contribute
\begin{eqnarray*}
\hat{H}_\uparrow &=& -\frac{1}{\beta}\log \int_0^{2\pi} d \theta_s \delta(\theta_s - \theta_{\ast})\\
&&
\hspace{10ex}
\times 
( e^{- \beta(2h - J)} \mid \bra{\psi_{s}}  \! \uparrow \rangle \mid^2 \\
&&
\hspace{10ex}+e^{\beta J} \frac{1}{2} \mid\bra{\psi_{s}}  \! \downarrow \rangle\mid^2
+e^{-\beta 3J} \frac{1}{2} \mid \bra{\psi_{s}}  \! \downarrow \rangle\mid^2
)
\\
&=&-\frac{1}{\beta}\log ( 
e^{- \beta(2h - J)} \cos^2{\theta_{\ast}} \\
& &\hspace{8ex} +e^{\beta J} \frac{1}{2} \sin^2{\theta_{\ast}}
+e^{-\beta 3J} \frac{1}{2}  \sin^2{\theta_{\ast}}
).
\end{eqnarray*}
\noindent Eigenpieces (d), (s3) and (a3) yield
\begin{eqnarray*}
\hat{H}_\downarrow &=& -\frac{1}{\beta}\log \int_0^{2\pi} d \theta_s \delta(\theta_s - \theta_{\ast})\\
&&
\hspace{10ex}
\times
( 
e^{- \beta(2h - J)} \mid \bra{\psi_{s}}  \! \downarrow \rangle \mid^2 \\
&&
\hspace{10ex}
+e^{\beta J} \frac{1}{2} \mid\bra{\psi_{s}}  \! \uparrow \rangle\mid^2
+e^{-\beta 3J} \frac{1}{2} \mid \bra{\psi_{s}}  \! \uparrow \rangle\mid^2
)
\\
&=&-\frac{1}{\beta}\log ( 
e^{- \beta(-2h - J)} \sin^2{\theta_{\ast}}\\
& &\hspace{8ex} 
+e^{\beta J} \frac{1}{2} \cos^2{\theta_{\ast}}
+e^{-\beta 3J} \frac{1}{2}  \cos^2{\theta_{\ast}}
).
\end{eqnarray*}

\subsection{The Gibbs free energy of the binary symmetric channel}
The inverse temperature times the system's Gibbs free energy 
\begin{equation}\label{gfe}
\beta \,\left( - \textrm{Tr}\{ \rho (\mathcal{\tilde{H}+\hat{H}- H})\} - \frac{1}{\beta}\log(Q) \right)
\end{equation}
measures the amount of free energy in the system that is not involved in the ``work of communication'', i.e. separating the
joint state into its products \cite{fordnote}. The energy eigenpiece contributions to the work calculation are gathered together in Table \ref{table:workfest}

\begin{table}[htpb] 
\caption{accumulated  eigenpiece contributions} % title of Table 
\centering      % used for centering table 
\begin{tabular}{c c c c c}  % centered columns (4 columns) 
\hline\hline  %inserts double horizontal lines 
\\  [-1.5ex]   %inserts vertical space before text starts 
label & \; $\tilde{H}$ & $\hat{H}$ & $H$ & $\rho$\\ [.5ex] % inserts table heading 
\hline   %inserts single horizontal line 
\\  [-1.5ex]                %inserts vertical space before text starts 
$(u)$ \; & \; ${\scriptstyle\tilde{H}_{\uparrow}\cos^2{\theta_{\ast}}}$ \; &\; ${\scriptstyle\hat{H}_{\uparrow}\cos^2{\theta_{\ast}}}$ &
 ${\scriptstyle(2h - J)\cos^2{\theta_{\ast}}}$ &  $e^{-\beta(2h-J)}$\\
\\
$(s1)$ \; & \; $\frac{\tilde{H}_{\uparrow}\cos^2{\theta_{\ast}}}{2}$ \; &
\; $\frac{\hat{H}_{\downarrow}\cos^2{\theta_{\ast}}}{2}$  & $\frac{-J\cos^2{\theta_{\ast}}}{2}$ &  $e^{\beta J}$\\
\\
$(s4)$ \; & \; $\frac{\tilde{H}_{\downarrow}\sin^2{\theta_{\ast}}}{2}$ \; &
\; $\frac{\hat{H}_{\uparrow}\sin^2{\theta_{\ast}}}{2}$  & $\frac{-J\sin^2{\theta_{\ast}}}{2}$ &  $e^{\beta J}$\\
\\
$(a1)$\; & \; $\frac{\tilde{H}_{\uparrow}\cos^2{\theta_{\ast}}}{2}$ \; &
\;  $\frac{\hat{H}_{\downarrow}\cos^2{\theta_{\ast}}}{2}$ & $\frac{3J \cos^2{\theta_{\ast}}}{2}$ &  $e^{-\beta 3J}$\\
\\
$(a4)$\; & \; $\frac{\tilde{H}_{\downarrow}\sin^2{\theta_{\ast}}}{2}$ \; &
\;  $\frac{\hat{H}_{\uparrow}\sin^2{\theta_{\ast}}}{2}$ & $\frac{3J \sin^2{\theta_{\ast}} }{2}$ &  $e^{-\beta 3J}$\\
\\
$(d)$\; & \; $\;{\scriptstyle \tilde{H}_{\downarrow}\sin^2{\theta_{\ast}}}$ \; & \; ${\scriptstyle \hat{H}_{\downarrow}\sin^2{\theta_{\ast}}}$ &
${\scriptstyle(-2h - J)\sin^2{\theta_{\ast}}}$ &  $e^{\beta(2h+J)}$\\ [1ex]       % [1ex] adds vertical space 
\hline     %inserts single line 
\end{tabular} 
\label{table:workfest}  % is used to refer this table in the text 
\end{table}

\begin{figure}[thtbp]
\begin{center}
\leavevmode
\includegraphics[width=60mm,keepaspectratio]{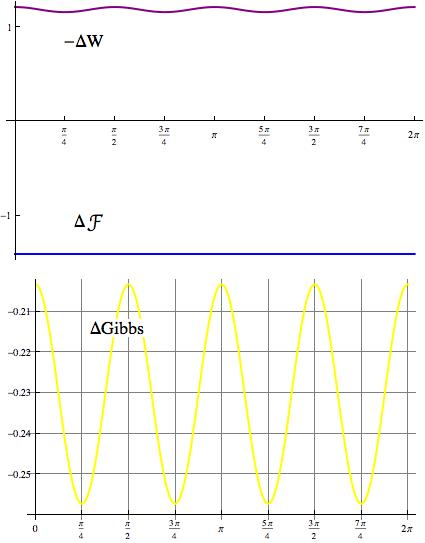}
\caption{ Upper frame: $- \beta \, \textrm{Tr}\{ \rho (\mathcal{\tilde{H}+\hat{H}- H})\}$, the work of separation and  $-\log(Q)$,
$\beta$ times the Helmholtz free energy of the system. Lower frame: their sum, the Gibbs free energy as a function of 
$\theta_s \in [0, 2\pi)$.}
\label{quantumgibbs}
\end{center}
\end{figure}

In figure \ref{quantumgibbs}, the system's Gibbs free energy 
\noindent is shown for a binary symmetric channel, governed by hamiltonian (\ref{h_hammy}),  
at parameter values $\beta=1$, $J=1$ and $h=0$. It is interesting to note that the Gibbs free energy
minimum for the quantum product state is higher than that of the analogous classical Ising system \cite{fordnote}.

\subsection{A comment on applications}
Figure \ref{wrinkle} demonstrates that for every choice of sender superposition state (the angle $\theta_s$), there are many possible 
receiver states (the angle $\theta_r$) accessible to the system. Yet, the performance of the quantum spin system under study has better Gibbs free energy properties than
the corresponding classical system \cite{fordnote}. 
In other words there is a trade off, the quantum analog
carries more information per bit but the sender does not have control over which bit is sent. One way to benefit from the
better communication properties of the quantum system and exercise a degree of control over the sent message is to collapse the wave
function on the sender's side of the channel and (depending upon the results of that measurement) apply a unitary transformation
to the system. The operators
$$
{\bf 1} \;
\begin{array}{ccc}
\ket{\uparrow}\\[1ex]
\ket{\downarrow}\
\end{array}
=
\begin{array}{ccc}
\ket{\uparrow}\\[1ex]
\ket{\downarrow}\
\end{array}
$$
\noindent corresponding to a phase shift $\theta \rightarrow \theta$ and
$$
{\bf X} \;
\begin{array}{ccc}
\ket{\uparrow}\\[1ex]
\ket{\downarrow}
\end{array}
=
\begin{array}{ccc}
\ket{\downarrow}\\[1ex]
\ket{\uparrow}\
\end{array}
$$
\noindent corresponding to a phase shift $\theta \rightarrow \theta + \frac{\pi}{2}$ are sufficient
to meet requirements of the present discussion.

Suppose that the sender intends to transmit the specific message $\uparrow$, $\downarrow$, $\uparrow$. 
The sender prepares three systems in states
$$
\begin{array}{lll}
\left( \ket{\psi_{send}} \otimes \ket{\psi_{rcv}} \right)_1\\[1ex]
\left( \ket{\psi_{send}} \otimes \ket{\psi_{rcv}} \right)_2\\[1ex]
\left( \ket{\psi_{send}} \otimes \ket{\psi_{rcv}} \right)_3.
\end{array}
$$
\noindent Upon measurement the sender may find the partially collapsed states
$$
\begin{array}{lll}
\left( \ket{\uparrow} \otimes \ket{\psi_{rcv}} \right)_1\\[1ex]
\left( \ket{\uparrow} \otimes \ket{\psi_{rcv}} \right)_2\\[1ex]
\left( \ket{\downarrow} \otimes \ket{\psi_{rcv}} \right)_3.
\end{array}
$$
There are ``errors'' in messages 2 and 3 in the sense that the sent message did not collapse to the intended
state. To remedy this, the sender applies the identity transformation ${\bf 1} \otimes {\bf 1}$ to the state $\left( \ket{\uparrow} \otimes \ket{\psi_{rcv}} \right)_1$ and the transformation ${\bf 1}\otimes{\bf X}$ to states 2 and 3.

\end{document}